\documentclass[conference]{IEEEtran}
\IEEEoverridecommandlockouts
\def\BibTeX{{\rm B\kern-.05em{\sc i\kern-.025em b}\kern-.08em
    T\kern-.1667em\lower.7ex\hbox{E}\kern-.125emX}}

\usepackage{subfigure} 
\usepackage[dvipdfmx]{graphicx} 
\graphicspath{{./image/}} 
\usepackage{booktabs}
\usepackage{multirow}
\usepackage{wrapfig}
\usepackage{amsmath}
\usepackage{amssymb}
\usepackage{amsfonts}
\usepackage{scalefnt}
\usepackage[colorlinks=true, allcolors=blue]{hyperref}
\usepackage{overcite}
\usepackage{cite}

\newcommand{\revise}[1]{\textcolor{black}{#1}}

\title{\LARGE 
Focal Surface Projection: Extending Projector Depth-of-Field Using a Phase-Only Spatial Light Modulator}

\author{\IEEEauthorblockN{Fumitaka Ueda\IEEEauthorrefmark{1},
Yuta Kageyama\IEEEauthorrefmark{2}, Daisuke Iwai\IEEEauthorrefmark{3} and
Kosuke Sato\IEEEauthorrefmark{4}}
\IEEEauthorblockA{Graduate School of Engineering Science, Osaka University\\
Email: \IEEEauthorrefmark{1}fumitaka.ueda@sens.sys.es.osaka-u.ac.jp,
\IEEEauthorrefmark{2}kageyama@sens.sys.es.osaka-u.ac.jp,\\
\IEEEauthorrefmark{3}daisuke.iwai@sys.es.osaka-u.ac.jp,
\IEEEauthorrefmark{4}sato@sys.es.osaka-u.ac.jp}}

\begin{document}

\maketitle
\thispagestyle{empty}
\pagestyle{empty}

\begin{abstract}
We present a focal surface projection to solve the narrow depth-of-field problem in projection mapping applications.
We apply a phase-only spatial light modulator to realize nonuniform focusing distances, whereby the projected contents appear focused on a surface with considerable depth variations.
The feasibility of the proposed technique was validated through a physical experiment.
\end{abstract}

\section{Introduction}\label{sec:introduction}

Projection mapping~(PM) is an augmented reality (AR) display technology in which computer-generated images are optically superimposed onto physical objects by projected imagery.
Researchers have found it useful in various fields, such as industrial design~\cite{8797923}, office work~\cite{10.1145/1959826.1959828, Iwai2011}, and education~\cite{hoang2017augmented}.
One of the major technical issues in practical applications of PM is defocus blur.
The lens aperture of a normal projector is designed to be sufficiently large to radiate bright images, which leads to a shallow projector's depth of field~(DoF).
Since a projection mapping target is a three-dimensional surface, a shallow DoF causes a spatially nonuniform defocus blur in the projected results.

Previous research aiming at extending the projector's DoF can be divided into single- and multiple-projector techniques.
Most single-projector techniques sharpen the original images such that the projected results look similar to the original ones.
This pre-sharpening is realized using the Wiener filter~\cite{brown2006image,oyamada2007focal}, constrained optimization~\cite{zhang2006projection,iwai2015extended}, or deep neural networks~\cite{kageyama2020prodebnet,kageyama2022online}.
Although these methods significantly reduce the defocus blur, the pre-sharpening process introduces ringing artifacts in the projected result when the point spread function (PSF) of a projected pixel cuts a large number of high-spatial-frequency components~\cite{grosse2010coded}.
Multiple-projector techniques place multiple projectors aiming at a target surface and select the one that can display the sharpest image for each point of the surface~\cite{bimber2006multifocal,nagase2011dynamic,10.1145/2508363.2508416,10.1145/2816795.2818111}.
Because there is no pre-sharpening process, they do not suffer from the ringing artifact problem.
However, preparing multiple projectors is costly, and the effective placement of the projectors is challenging.

We propose a single-projector technique that adopts a \textit{focal surface} approach to extend a projector's DoF.
Focal surfaces constitute a recently explored optical design that represents spatially varying focal lengths of an optical system.
For example, a previous \revise{publication} placed the virtual image of a displayed three-dimensional (3D) scene with spatially nonuniform depths in a virtual reality head-mounted display (HMD) at the desired distances from a user's viewpoint~\cite{matsuda2017focal}.
Another work proposed a technique that places the virtual image of an occlusion mask just behind a 3D virtual object displayed in an optical see-through HMD~\cite{Hiroi:21}.
These works used phase-only spatial light modulators (PSLM) as controllable free-form lenses to realize the focal surfaces.
In this paper, we develop a novel projection system that applies a PSLM as its objective lens, and we propose a computational technique that optimizes the spatial phase pattern of the PSLM so that the projected results appear focused on a target surface with spatially varying depths from the projector.
Our computational technique deals with the real image of the projector's display panel, while the above-mentioned works deal with virtual images.
We build a physical prototype and conduct an experiment to confirm that the proposed technique alleviates the defocus blur problem in PM for a surface with spatially non-constant depths from the projector.

\section{Principles of Focal Surface Projection}\label{sec:principle}

\begin{figure}[t]
\centering\includegraphics[width=0.99\linewidth]{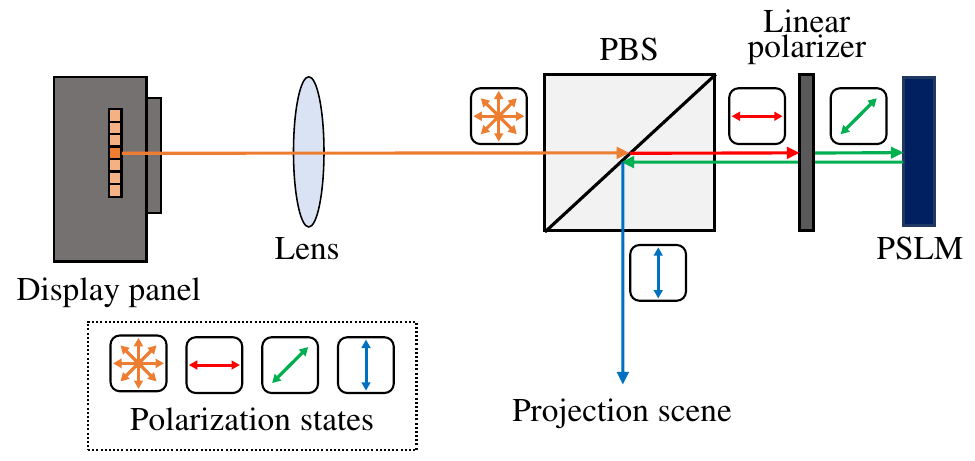}
\caption{Schematic diagram of the proposed system.}
\label{fig:proposed_system}
\end{figure}


We aimed to realize a DoF extension of a projector by optically bending its focal plane to fit a projection surface.
%
%
\autoref{fig:proposed_system} shows our optical design to achieve this goal.
Light emitted from the projector's display panel passes through a convex lens, a polarization beam splitter (PBS), and a linear polarizer and is then incident on a PSLM.
After the phase of the incident light is modulated on the PSLM, the light travels back through the linear polarizer, is reflected by the PBS, and is finally superimposed onto a projection surface.
A typical PSLM is implemented by a liquid crystal on silicon (LCoS) that modulates the phase of the incident light only when its \revise{polarization} direction is in a particular direction, which we suppose to be 45 degrees in this paper without loss of generality.
In \autoref{fig:proposed_system}, the \revise{polarization} direction of the projected light becomes 0 and 45 degrees after it passes through the PBS and linear polarizer, respectively.   
The reflected light on the PBS then passes through the same linear polarizer, whereby the \revise{polarization} direction is rotated to 90 degrees.
Finally, the light is reflected in the PBS toward the projection surface.


An LCoS-based PSLM accomplishes the phase delay by electrically adjusting the optical refractive index along the light path~\cite{Zhang2014}.
We can control the refractive index of each LCoS pixel using a grayscale image, which is called a phase image.
The rest of this section describes how the phase image is computed given the depth of a target surface from the PSLM of the proposed projector.

\begin{wrapfigure}{r}{4.0cm}
\centering
	\includegraphics[width=0.98\hsize]{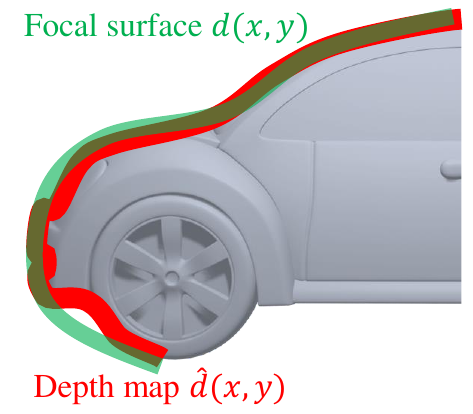}
	\caption{Focal surface approximation of a spatially non-constant depth map.}
	\label{fig:focal_surface}
\end{wrapfigure}
First, we approximate the depth map of a target surface to a smooth focal surface.
Suppose that $(x,y)$ is the coordinate value of a projector pixel that is incident on a point of the target surface (\autoref{fig:focal_surface}).
Suppose, further, that $\hat{d}(x,y)$ and $d(x,y)$ represent the depth value of the point and its approximated depth value on the focal surface, respectively.
We compute this approximation by solving the following optimization problem:
%
%
\begin{equation}\label{equ:3}
    \min_{d}{\sum_{(x,y)}\|\hat{d}(x,y)-d(x,y)\|^2 + \gamma\sum_{(x,y)}\|\partial^2 d(x,y)\|^2}.
\end{equation}
Note that the second term in the optimization problem is necessary to minimize optical aberrations~\cite{matsuda2017focal}, and $\gamma$ is a weight parameter.
We use the nonlinear least squares method to solve \autoref{equ:3}.


Suppose that a ray from the projector pixel $(x,y)$ passes through the optical center of the first lens and is incident on the PSLM at its $(p_x,p_y)$ coordinate.
We optimize the local phase pattern of the PSLM around $(p_x,p_y)$ to realize a target focal length $f_p(p_x,p_y)$ whereby the rays emitted from the pixel $(x,y)$ focus on the corresponding point of the focal surface $d(x,y)$.
In the next step, we determine the target focal length using a ray transfer matrix~(RTM).

\begin{figure}[t]
\centering\includegraphics[width=0.99\linewidth]{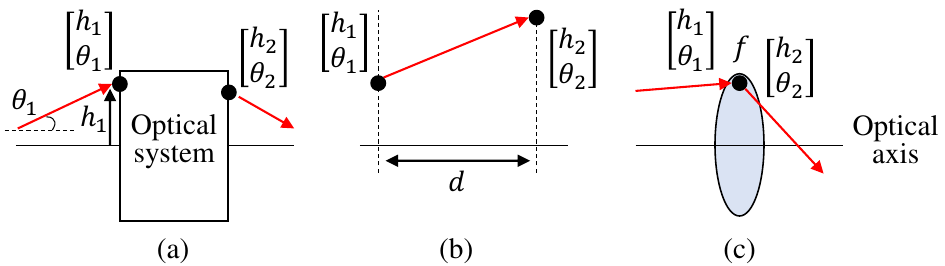}
\caption{RTM representation of rays in optical systems.}
\label{fig:Ray_Transfer}
\end{figure}

In an RTM, a ray is represented by height $h$ and angle $\theta$ from the optical axis.
We consider a ray~$(h_1, \theta_1)$ that passes through an optical system and enters a new state~$(h_2,\theta_2)$~(\autoref{fig:Ray_Transfer}(a)).
The relationship can be expressed by the following equation using the RTM~\cite{ersoy2006diffraction}:
\begin{equation}\label{equ:4}
    \begin{bmatrix}
       h_2  \\
       \theta_2 
    \end{bmatrix}
    = 
    \begin{bmatrix}
       A & B  \\
       C & D 
    \end{bmatrix}
    \begin{bmatrix}
       h_1  \\
       \theta_1 
    \end{bmatrix}.
\end{equation}
When a ray travels a distance $d$ along the optical axis through a medium with a constant refractive index, the RTM $\mathbf{P}(d)\in\mathbb{R}^{2\times2}$ takes the values of $(A,B,C,D)=(1,d,0,1)$~(\autoref{fig:Ray_Transfer}(b)).
When a ray passes through a thin lens with a focal length $f$, the RTM $\mathbf{L}(f)\in\mathbb{R}^{2\times2}$ takes the values of $(A,B,C,D)=(1,0,-1/f,1)$~(\autoref{fig:Ray_Transfer}(c)).
When an optical system consists of multiple optical elements, the RTM of the entire system is simply derived from the product of the RTM of each element. 

\begin{figure}[t]
\centering\includegraphics[width=0.99\linewidth]{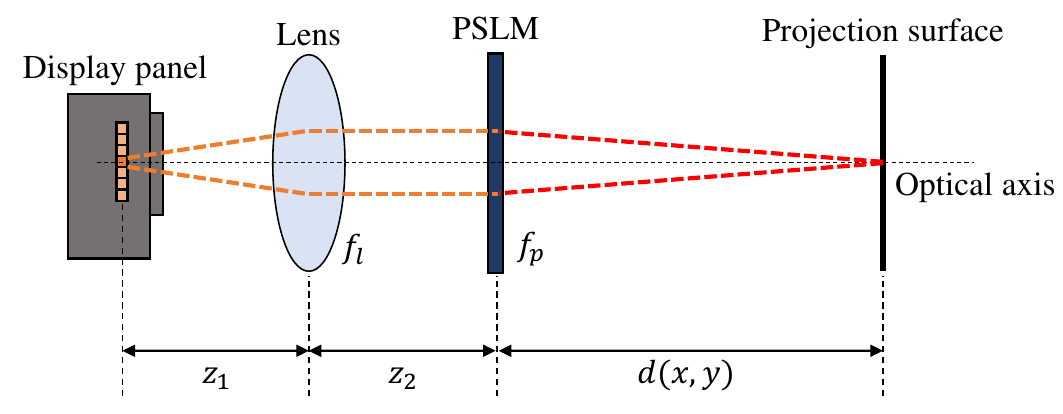}
\caption{Effective location of each optical component of the proposed system along the optical axis.}
\label{fig:RTM}
\end{figure}

In our setup, the convex lens with the focal length of $f_l$ is placed at a distance $z_1$ from the projector's display panel (\autoref{fig:RTM}).
The PSLM with the target focal length of $f_p(p_x,p_y)$ is placed at a distance $z_2$ from the lens.
The distance between the PSLM and the focal surface is $d(x,y)$.
The RTM $\mathbf{M}(x,y)\in\mathbb{R}^{2\times2}$ of this setup then becomes
\begin{equation}\label{equ:7}
    \mathbf{M}(x,y) = \mathbf{P}(d(x,y))\mathbf{L}(f_p(p_x,p_y))\mathbf{P}(z_2)\mathbf{L}(f_l)\mathbf{P}(z_1).
\end{equation}
The $(1,2)$ element of $\mathbf{M}(x,y)$ must be zero so that all the rays emitted from the projector pixel $(x,y)$ converge at the corresponding focal surface point $d(x,y)$.
Then, we obtain the target focal length of the PSLM:
\begin{equation}\label{equ:8}
    \hspace{-6pt}
    f_p(p_x,p_y) = \frac{\{f_l z_1 + f_l z_2 - z_1 z_2\}d(x,y)}{f_l z_1 + f_l z_2 - z_1 z_2 + f_l d(x,y) - z_1 d(x,y)}.
\end{equation}


In the final step, we compute the phase image $\phi$. 
To achieve a focal length of $f_p(p_x,p_y)$ with the neighboring pixels of the PSLM, the Hessian matrix of the phase image at $(p_x,p_y)$, denoted as $\mathbf{H}_{\phi}(p_x, p_y)\in\mathbb{R}^{2\times2}$, must satisfy the following equation~\cite{matsuda2017focal}:
%
\begin{equation}\label{equ:9}
    \mathbf{H}_{\phi}(p_x, p_y) = -\frac{2\pi}{\lambda f_{p}(p_x,p_y)}\mathbf{I},
\end{equation}
where $\mathbf{I}\in\mathbb{R}^{2\times2}$ is an identity matrix, and $\lambda$ is the wavelength of the incident light.
However, a $\phi$ that satisfies this equation exists only when $f_p$ is spatially constant, in which the PSLM functions as a lens with a single focal length.
Therefore, we solve the following linear least squares problem to obtain a $\phi$ with a Hessian matrix as close as possible to the Hessian matrix that satisfies \autoref{equ:9}:
\begin{equation}\label{equ:10}
    \min_{\phi}{\sum_{(p_x, p_y)}\|\hat{\mathbf{H}}_{[\phi]}(p_x,p_y)-\frac{-2\pi}{\lambda f_p(p_x, p_y)}\mathbf{I}\|^{2}_{F}},
\end{equation}
where $\|\cdot\|^{2}_{F}$ is the Frobenius norm, and $\hat{\mathbf{H}}_{[\cdot]}$ is the discrete Hessian operator.



\begin{figure}[t]
\centering\includegraphics[width=0.99\linewidth]{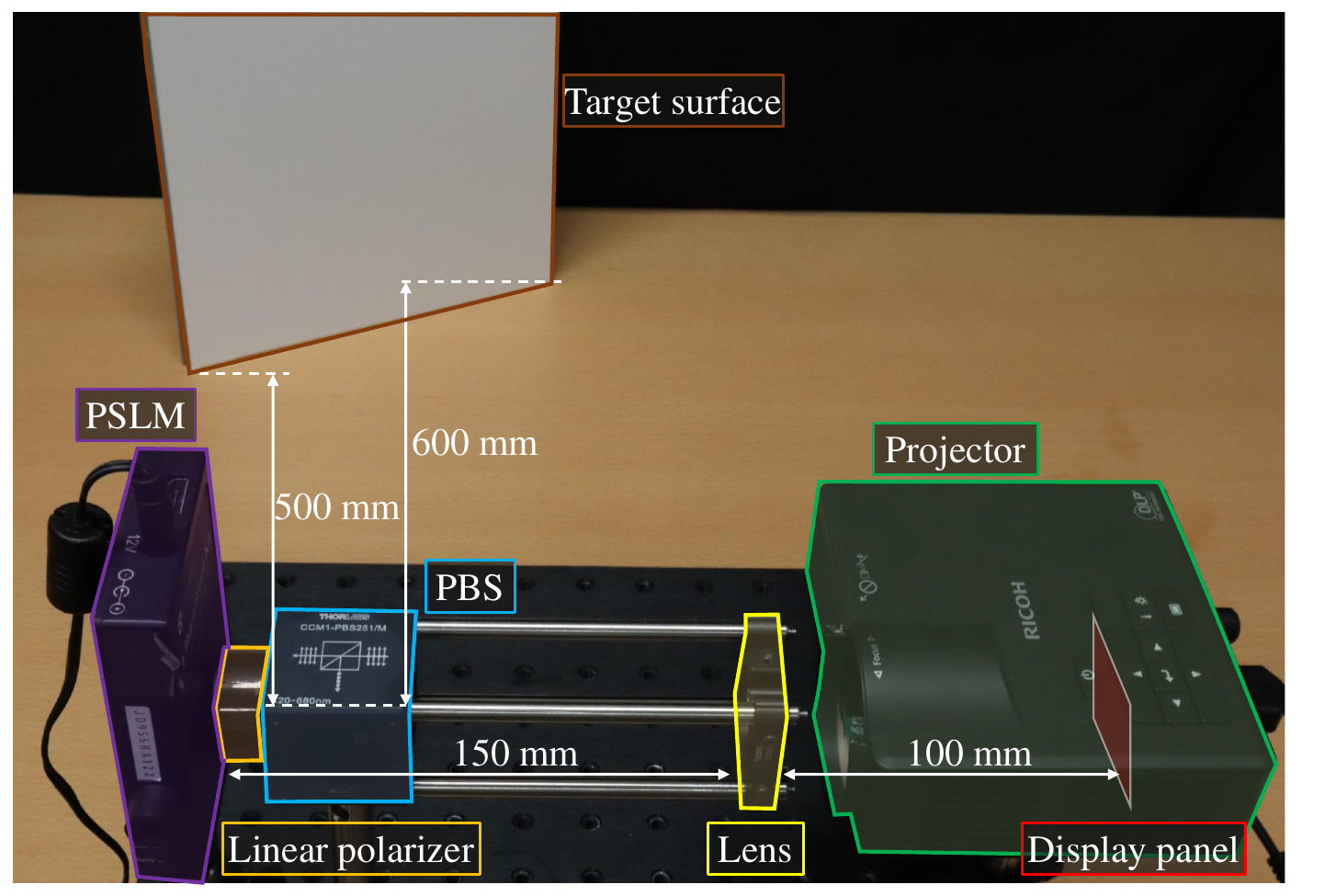}
\caption{The prototype system and target surface in the experiment.}
\label{fig:prototype}
\end{figure}

\begin{figure}[t]
\centering\includegraphics[width=0.95\linewidth]{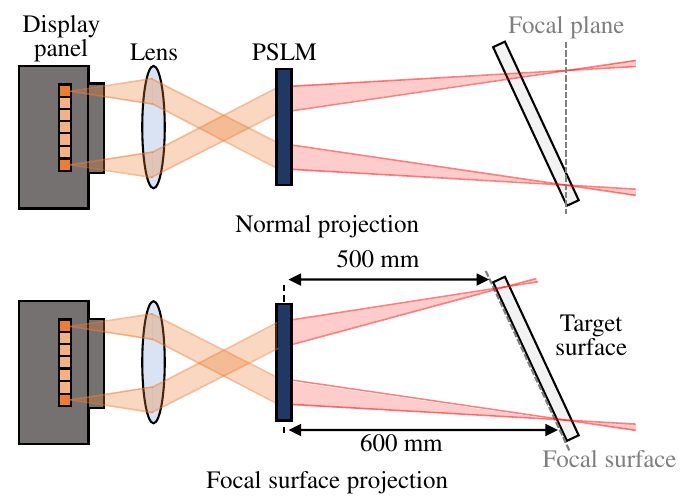}
\caption{Convergence of projected rays in the normal and focal surface projections.}
\label{fig:condition}
\end{figure}

\begin{figure}[t]
\vspace{-10pt}
\centering\includegraphics[width=0.99\linewidth]{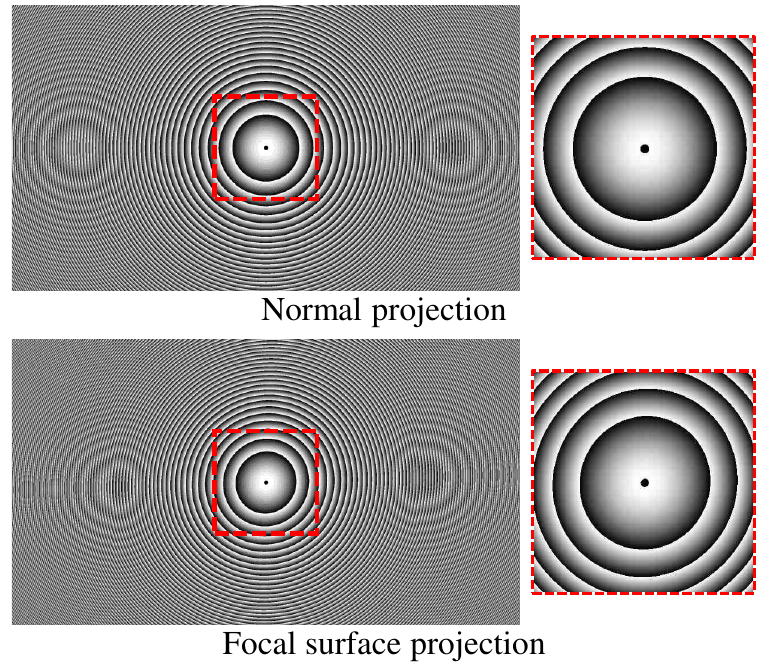}
\caption{Phase images computed for the normal and focal surface projections.}
\label{fig:phase_image}
\end{figure}

\begin{figure}[t]
\centering\includegraphics[width=0.99\linewidth]{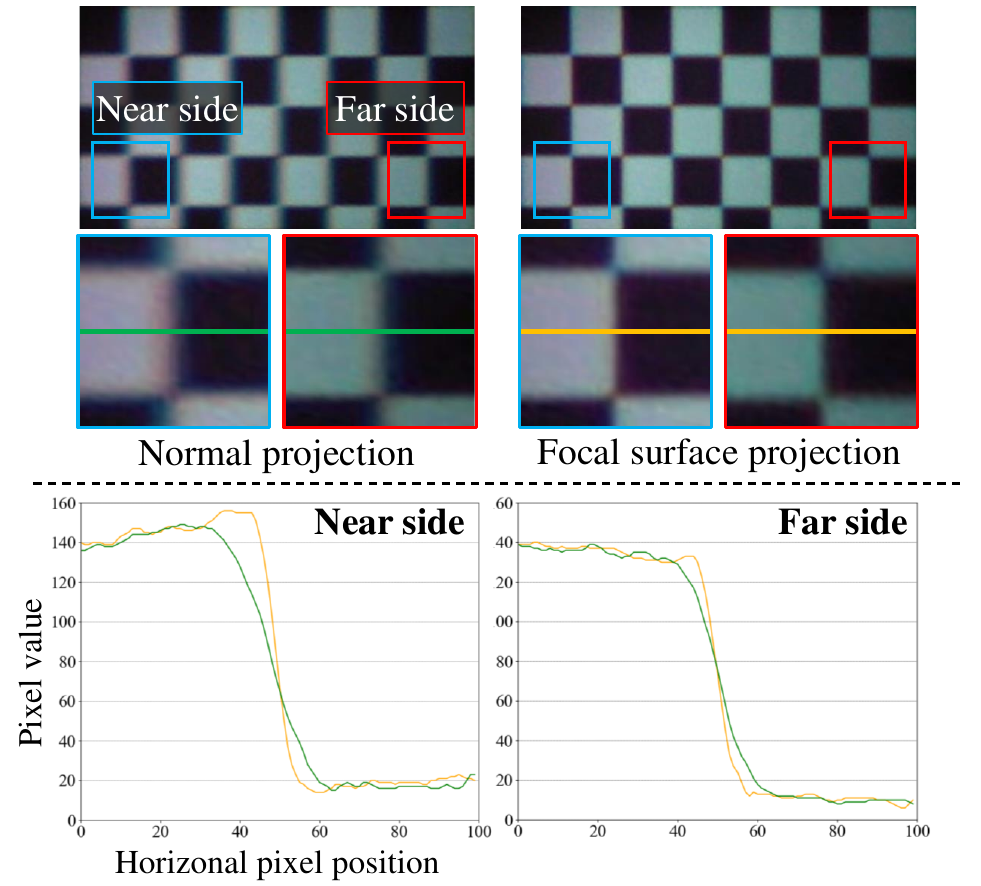}
\caption{Projected chess patterns in the normal projection and the focal surface projection conditions. The graphs show the intensity profiles along the green (normal projection) and yellow (focal surface projection) lines. The target intensity pattern is a simple step function.}
\label{fig:chess}
\end{figure}

\begin{figure}[t]
\centering\includegraphics[width=0.99\linewidth]{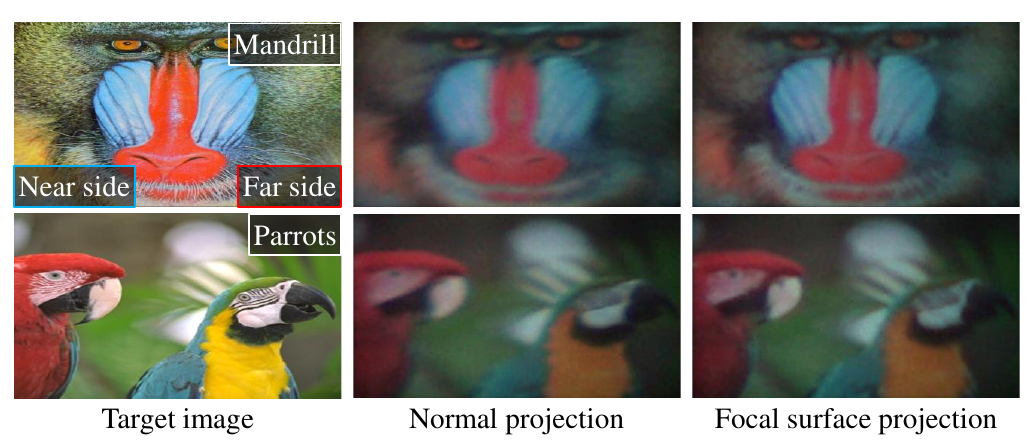}
\caption{Projected results of the natural images.}
\label{fig:result}
\end{figure}

\section{Experiment}

%
%
We built a prototype system, shown in \autoref{fig:prototype}, and conducted an experiment to validate the proposed focal surface projection technique.
We used a digital light processing projector (PJ WXC1110; RICOH), from which the objective lens was removed as the display panel and the light source of the system.
The other optical elements were a convex lens~(AC254-100-A; Thorlabs), a PBS~(CCM1-PBS251/M; Thorlabs), a linear polarizing plate~(LPVISE100-A; Thorlabs), and a PSLM~(JD955B EDucation Kit; Jasper Display).
When solving \autoref{equ:9} and \ref{equ:10}, we used a fixed $\lambda$ value of 532 nm that corresponds to the wavelength of the projector's green channel.
We prepared two experimental conditions: \emph{focal surface projection} and \emph{normal projection}. 


In front of the system, we placed a slanted planar surface, by which the depth from the PSLM to the surface became spatially non-constant (\autoref{fig:condition}).
The depth range of the surface ranged from 500 mm to 600 mm.
In the focal surface projection condition, we computed the phase image for this target surface using the method described in Section \ref{sec:principle}.
In the normal projection condition, we prepared another phase image that represented a simple convex lens whereby the system behaved as a normal projector, whose focusing distance was 600 mm from the PSLM.
The phase images in the two conditions are shown in \autoref{fig:phase_image}.

\autoref{fig:chess} shows the projected results of a chess pattern under the two conditions, which were captured by a camera perfectly focused on the entire surface area.
In the normal projection condition, the chess edges appear blurred around the near side of the target surface.
In the focal surface projection condition, the edges appear less blurred.
\autoref{fig:chess}, which shows the intensity profiles along the edges, confirms the improvement quantitatively.

\autoref{fig:result} shows the projected results of two natural images, which were captured by the same camera.
Like the projected chess pattern images, the projected natural images appear blurred around the near side of the target surface in the normal projection condition, while they appear less blurred in the proposed focal surface projection condition.
We quantitatively evaluated the image quality of the projected results using the standard metrics: structural similarity~(SSIM)~\cite{wang2004image} and learned perceptual image patch similarity~(LPIPS)~\cite{zhang2018unreasonable}.
We did not use the peak signal-to-noise ratio~(PSNR) because it is difficult to accurately trim the projected region from the captured image.
On the contrary, SSIM and LPIPS are robust for the misalignment between the reference and the evaluated images.
\autoref{tab:comparison} shows the results, which confirm that the proposed method can display images of better quality than the normal projection system.
\revise{In addition, since the difference was slight, we conducted a user experiment to clarify further the difference between the proposed method and the normal projection.
We asked the 12 participants (10 males and two females, aged 21 to 25), 'Which one appears sharper; the proposed method or the normal projection?'.
As a result, all participants answered that the proposed method appeared sharper.}

\begin{table}[t]
\centering
\caption{Quantitative comparison of normal projection and focal surface projection. \revise{Note that a higher SSIM and a lower LPIPS indicate better quality.}}
\label{tab:comparison}
\scalebox{1}{

\begin{tabular}{c|cc|cc}
\multirow{2}{*}{Metric} & \multicolumn{2}{c|}{Normal projection}& \multicolumn{2}{c}{Focal surface projection}\\ \cline{2-5} 
                         & Mandrill       & Parrots        & Mandrill       & Parrots        \\ \hline
SSIM ($\uparrow$)                    & 0.291          & 0.657 & \textbf{0.295} & \textbf{0.660}          \\
LPIPS ($\downarrow$)                 & 0.859          & 0.546 & \textbf{0.852} & \textbf{0.531}          
\end{tabular}
}
\end{table}


Nevertheless, although our focal surface projection system improved the quality of the projected images in comparison with the na\"{i}ve projection system, the defocus blur was still not perfectly removed.
We believe that there are several reasons for this.
One reason is the imperfection of the applied optics.
Another reason is that we supposed a fixed $\lambda$ when optimizing the phase image.
Therefore, the PSLM with a phase image optimized for a single-color channel may have caused chromatic aberrations in the projected result.
A solution to this issue would be to use a PSLM with a refresh rate of 180 Hz or higher and modulate it with different phase images optimized for each color channel.
Additionally, pre-sharpening methods such as that used in previous \revise{publications}~\cite{kageyama2020prodebnet, oyamada2007focal} could be used to further improve the quality of the projected images.
Because in the proposed method, the PSF does not significantly degrade the high-spatial-frequency components, pre-sharpening should not cause significant ringing artifacts.


\section{Conclusion}

We presented a focal surface projection technique to extend a projector's DoF for PM applications.
Our main contribution is the proposed phase image computation method, which achieves spatially varying focusing distances of a single projector.
We conducted an experiment using a prototype system and confirmed that our focal surface projection could display images of better quality, with less defocus blur, on a slanted surface than a normal projection system.
In a future study, we aim to extend our computational framework to support dynamic PM applications.

\bibliography{bibliography} 

\begin{thebibliography}{10}

\bibitem{8797923}
Takezawa T, Iwai D, Sato K, Hara T, Takeda Y, and Murase K.
\newblock Material surface reproduction and perceptual deformation with
  projection mapping for car interior design.
\newblock In {\em Proceedings of the IEEE Conference on Virtual Reality and 3D
  User Interfaces (VR)}, 2019;251--258.

\bibitem{10.1145/1959826.1959828}
Matsushita K, Iwai D, and Sato K.
\newblock Interactive bookshelf surface for in situ book searching and storing
  support.
\newblock In {\em Proceedings of the 2nd Augmented Human International
  Conference}, 2011;1--8.

\bibitem{Iwai2011}
Iwai D and Sato K.
\newblock Document search support by making physical documents transparent in
  projection-based mixed reality.
\newblock {\em Virtual Reality}, 2011;15(2):147--160.

\bibitem{hoang2017augmented}
Hoang T, Reinoso M, Joukhadar Z, Vetere F, and Kelly D.
\newblock Augmented studio: Projection mapping on moving body for physiotherapy
  education.
\newblock In {\em Proceedings of the 2017 CHI conference on human factors in
  computing systems}, 2017;1419--1430.

\bibitem{brown2006image}
Brown MS, Song P, and Cham T.
\newblock Image pre-conditioning for out-of-focus projector blur.
\newblock In {\em Proceedings of IEEE Conference on Computer Vision and Pattern
  Recognition}, 2006;1956--1963.

\bibitem{oyamada2007focal}
Oyamada Y and Saito H.
\newblock Focal pre-correction of projected image for deblurring screen image.
\newblock In {\em Proceedings of IEEE Conference on Computer Vision and Pattern
  Recognition}, 2007;1--8.

\bibitem{zhang2006projection}
Zhang L and Nayar S.
\newblock Projection defocus analysis for scene capture and image display.
\newblock {\em ACM Trans. Graph.}, 2006;25(3):907--915.

\bibitem{iwai2015extended}
Iwai D, Mihara S, and Sato K.
\newblock Extended depth-of-field projector by fast focal sweep projection.
\newblock {\em IEEE Trans. Vis. Comput. Graph.}, 2015;21(4):462--470.

\bibitem{kageyama2020prodebnet}
Kageyama Y, Isogawa M, Iwai D, and Sato K.
\newblock Prodebnet: projector deblurring using a convolutional neural network.
\newblock {\em Opt. Express}, 2020;28(14):20391--20403.

\bibitem{kageyama2022online}
Kageyama Y, Iwai D, and Sato K.
\newblock Online projector deblurring using a convolutional neural network.
\newblock {\em IEEE Trans. Vis. Comput. Graph.}, 2022;28(5):2223--2233.

\bibitem{grosse2010coded}
Grosse M, Wetzstein G, Grundh{\"o}fer A, and Bimber O.
\newblock Coded aperture projection.
\newblock {\em ACM Trans. Graph.}, 2010;29(3):1--12.

\bibitem{bimber2006multifocal}
Bimber O and Emmerling A.
\newblock Multifocal projection: A multiprojector technique for increasing
  focal depth.
\newblock {\em IEEE Trans. Vis. Comput. Graph.}, 2006;12(4):658--667.

\bibitem{nagase2011dynamic}
Nagase M, Iwai D, and Sato K.
\newblock Dynamic defocus and occlusion compensation of projected imagery by
  model-based optimal projector selection in multi-projection environment.
\newblock {\em Virtual Reality}, 2011;15(2):119--132.

\bibitem{10.1145/2508363.2508416}
Bermano A, Br\"{u}schweiler P, Grundh\"{o}fer A, Iwai D, Bickel B, and Gross M.
\newblock Augmenting physical avatars using projector-based illumination.
\newblock {\em ACM Trans. Graph.}, 2013;32(6):1--10.

\bibitem{10.1145/2816795.2818111}
Siegl C, Colaianni M, Thies L, Thies J, Zollh\"{o}fer M, Izadi S, Stamminger M,
  and Bauer F.
\newblock Real-time pixel luminance optimization for dynamic multi-projection
  mapping.
\newblock {\em ACM Trans. Graph.}, 2015;34(6):1--11.

\bibitem{matsuda2017focal}
Matsuda N, Fix A, and Lanman D.
\newblock Focal surface displays.
\newblock {\em ACM Trans. Graph.}, 2017;36(4):1--14.

\bibitem{Hiroi:21}
Hiroi Y, Kaminokado T, Ono S, and Itoh Y.
\newblock Focal surface occlusion.
\newblock {\em Opt. Express}, 2021;29(22):36581--36597.

\bibitem{Zhang2014}
Zhang Z, You Z, and Chu D.
\newblock Fundamentals of phase-only liquid crystal on silicon (lcos) devices.
\newblock {\em Light Sci. Appl.}, 2014;3(10):e213.

\bibitem{ersoy2006diffraction}
Ersoy OK.
\newblock {\em Diffraction, Fourier optics and imaging}.
\newblock John Wiley \& Sons, 2006.

\bibitem{wang2004image}
Wang Z, Bovik AC, Sheikh HR, and Simoncelli EP.
\newblock Image quality assessment: from error visibility to structural
  similarity.
\newblock {\em IEEE Trans. Image Process.}, 2004;13(4):600--612.

\bibitem{zhang2018unreasonable}
Zhang R, Isola P, Efros AA, Shechtman E, and Wang O.
\newblock The unreasonable effectiveness of deep features as a perceptual
  metric.
\newblock In {\em Proceedings of IEEE Conference on Computer Vision and Pattern
  Recognition}, 2018;586--595.

\end{thebibliography}
\bibliographystyle{junsrt} 

\end{document}